\documentclass[aps,twocolumn,superscriptaddress]{revtex4-2}

\usepackage{amsmath,amsfonts,amssymb}

\usepackage{mathabx} 
\usepackage{color}
\usepackage{graphicx}
\usepackage{physics}
\usepackage[colorlinks,citecolor=green,urlcolor=blue,bookmarks=false,hypertexnames=true]{hyperref} 

\begin{document}

\title{A Gauss-Bonnet Theorem for Quantum States: Gauss Curvature and Topology in the Projective Hilbert Space}
\author{Shin-Ming Huang}
\affiliation{Department of Physics, National Sun Yat-sen University, Kaohsiung 80424, Taiwan}
\affiliation{Center for Theoretical and Computational Physics, National Sun Yat-sen University, Kaohsiung 80424, Taiwan}
\affiliation{Physics Division, National Center for Theoretical Sciences, Taipei 10617, Taiwan}

\begin{abstract}
Geometry and topology are fundamental to modern condensed matter physics, but their precise connection in quantum systems remains incompletely understood. Here, we develop an analytical scheme for calculating the curvature of the quantum metric of Bloch bands. Using a gauge-invariant formulation based on eigenprojectors, we construct the full Riemannian geometry of the quantum-state manifold and apply it to a two-dimensional two-band model. We find that the Gauss curvature is constant over regular regions, but the manifold inevitably develops a closed curve of singular points where the metric tensor degenerates. These singularities obstruct the conventional Gauss–Bonnet theorem. By introducing the notion of a front and a signed area form, we derive a generalized Gauss–Bonnet relation that includes a singular curvature term defined along the fold curve. This result establishes a direct, quantized link between the total signed Gauss curvature and the Chern number, providing a unified geometric interpretation of Berry curvature and quantum metric. This framework bridges differential geometry and topological band theory, revealing how singular folds mediate the discrepancy between quantum volume and topological charge.
\end{abstract}

\maketitle

\section{Introduction}
The geometric structure of quantum states encodes profound information about the physical properties and topology of quantum systems. In crystalline solids, the local geometry of Bloch states is captured by the quantum geometric tensor (QGT), whose real and imaginary parts correspond respectively to the quantum metric~\cite{provost1980} and the Berry curvature~\cite{berry1984}. The Berry curvature has been widely studied as the source of geometric phases and topological invariants such as the Chern number, that underlie quantized Hall responses and topological insulators \cite{thouless1982,hasan2010,qi2011}. 
Most previous studies have focused on the Berry curvature because the quantum world is inherently complex~\cite{page1987}; the Hilbert space of states forms a complex vector bundle, and its geometric properties are naturally described by a connection with a U(1) gauge field. This view has driven decades of research in topological phases, where the complex structure of the quantum bundle directly gives rise to quantized observables \cite{niu1985, avron1985,kohmoto1985}. 

In contrast, the real part of the QGT—the quantum metric—was long considered secondary. Only recently has it been recognized that the metric plays an equally fundamental role~\cite{gao2023quantum,wang2023quantum,qi2025,liu2024,cuerda2024,yu2025quantum,gao2025quantum,liu2025quantum}, characterizing the local Riemannian geometry of the quantum-state manifold and affecting physical quantities. At first, the quantum metric was shown to relate to Wannier localization~\cite{resta1999,marzari1997,resta2011insulating}, superfluid stiffness \cite{peotta2015}, flat-band superconductivity \cite{torma2022}, and orbital susceptibility \cite{roy2014band}. Recent studies show the quantum metric is associated to longitudinal conductivity~\cite{huhtinen2023,qi2025}, nolinear Hall responses in antiferromagnetic heterostructures \cite{gao2023quantum,wang2023quantum}, and other higher-order responses~\cite{ahn2022riemannian,fang2024quantum}.

Efforts to explore the Riemannian structure of quantum systems have revealed deep parallels between quantum geometry and classical differential geometry. The quantum metric defines a natural notion of distance between nearby Bloch states \cite{provost1980, peotta2015}, and its curvature can, in principle, encode global topological information. Several studies have examined how the quantum metric and the Berry curvature jointly determine the quantum volume \cite{monaco2017, mera2021, mera2022nontrivial}, suggesting an underlying curvature–topology correspondence reminiscent of the Gauss–Bonnet theorem in classical geometry \cite{do2016differential}. However, this correspondence remains incomplete. Recent attempts to unify quantum and classical geometric formulations \cite{ahn2022riemannian, hetenyi2023fluctuations, chen2024quantum, wang2024geodesic} have not yet yielded a fully consistent framework. Motivated by this gap, the author has examined quantum geometry through the information of quantum distances between eigenstate projectors (eigenprojectors), revealing that these projectors can be naturally interpreted as coordinates in a Euclidean space \cite{huang2024exploring}. This insight prompts a deeper investigation into the geometry of the manifold formed by eigenprojectors—specifically, a Grassmannian—and its associated topological features.

In this work, we derive the metric tensor, Christoffel symbols, and Riemann curvature directly from the eigenprojector of a quantum state. Since the eigenprojector can be expressed as a polynomial of the Bloch Hamiltonian \cite{graf2021}, we provides a general scheme for computing geometric quantities in arbitrary multi-band systems. We apply this framework to a two-band model and demonstrate that the quantum manifold exhibits a closed curve of singular points where the metric degenerates. These singularities manifest as folds and cusps in the quantum-state manifold, where conventional differential-geometric descriptions break down. 
To address this challenge, we adopt the concept of a front, which ensures an immersive map when a well-defined normal vector field exists. We introduce a singular curvature along the degeneracy curve and formulate a generalized Gauss–Bonnet theorem that remains valid even in the presence of these singularities. This leads to a striking result: the total signed Gauss curvature is directly related to the Chern number, implying that the Berry curvature is exactly the signed area density of the quantum manifold.
Our findings establish a unified differential-geometric framework that elucidates the interplay between quantum metric, Berry curvature, and topological invariants, offering new insights into the geometric foundations of quantum systems.

\section{Quantum Geometry}
\subsection{\label{sec:level1} Eigenprojectors}
Assume a Bloch Hamiltonian at crystal momentum $\vb{k}$
\begin{equation}
    H(\vb{k}) = \sum_{\alpha=1}^{N} E_\alpha (\vb{k}) \hat{P}_\alpha (\vb{k}),
\end{equation}
where $E_\alpha (\vb{k})$ is the eigenenergy and $\hat{P}_\alpha (\vb{k})= \ketbra{u_\alpha (\vb{k})}{u_\alpha (\vb{k})}$ is the eigenprojector with the band index $\alpha$ and crystal momentum $\vb{k}=(k_x,k_y)$. The eigenprojector for different bands at the same $\vb{k}$ are orthogonal to each other, and they are idempotent, giving $ \hat{P}_\alpha^2 = \hat{P}_\alpha$. From now on, we will omit the argument $\vb{k}$ unless it is specified.
In our previous work, we demonstrated that the eigenprojector can be interpreted as coordinates of a surface in Euclidean space, and the surface is suggested to be the manifold of the quantum system. A quick understanding is that by vectorizing the eigenprojector as $\vec{X}_\alpha =(\Re (\hat{P}_\alpha)_{11}, \Re (\hat{P}_\alpha)_{12},\cdots,\Im (\hat{P}_\alpha)_{11}, \Im (\hat{P}_\alpha)_{12}, \cdots)$, the trace form,  $\trace(\hat{P}_{\alpha} \hat{P}_{\beta}) = \sum_{ij} (\hat{P}_\alpha)_{ij}^* (\hat{P}_\beta)_{ij} = \vec{X}_\alpha \cdot \vec{X}_\beta$, is interpreted as a scalar product of two real vectors. (We let $\vec{X}_\alpha$ live in an ambient Euclidean space.)
In this representation, the vector has $2 N^2$ components, but some are redundant; for example, $\Im (\hat{P}_\alpha)_{ii}=0$. Only $N^2$ components are required using Hermitian basis matrices. 

The great advantage of utilizing the eigenprojectors is that they are gauge-invariant, so are their derivatives and the subsequent trace form. It is also true for the eigenprojectors containing many valence bands, $\hat{P}=\sum_{\alpha\in \mathrm{val.}} \hat{P}_\alpha$ (or bands of interest, especially for bands that are not separable). For single-state projectors, their trace form is the absolute value of their wavefunction overlap. For multi-state projectors, their trace form is the sum of the squared singular values of their overlap. This practical property of getting rid of gauge redundancy is the characteristic of the Grassmannian that identifies systems related by gauge conjugation as projecting them into a conjugacy class.

\subsection{Riemann curvature}
With this sense $\vec{X} \dot= \hat{P}$, we can study the geometry of the manifold. We need to investigate the change of the tangent planes of the surface from point to point. The tangent basis vectors are naturally defined as $\vec{e}_\mu \dot = \partial_{\mu}\hat{P} $ ($ \partial_{\mu}:= \frac{\partial}{\partial k^{\mu}} $ and $\mu=1,2$ for $x,y$, respectively, and $k^{\mu}=k_{\mu}$ throughout the paper.). The tangent basis vectors are the differential maps (images) of the basis vectors in the domain ($k$ space). Since the domain is two-dimensional, the tangent plane and then the quantum manifold is two-dimensional. 
  
The metric tensor of the manifold is defined as the inner products of basis vetors:
\begin{equation}
    g_{\mu \nu} = \vec{e}_\mu \cdot \vec{e}_\nu = \trace \left[ (\partial_{\mu}\hat{P}) (\partial_{\nu}\hat{P}) \right]. \label{metric}
\end{equation}
(We note that our definition of the metric tensor is twice the value of the conventional one.) The lengths might depend on the parametrization of $\vb{k}$. To obtain quantities that are independent of the choice of parameters, a dual coordinate system is introduced that is contravariant to compensate for each other. The dual basis vectors are denoted with an upper index: $\vec{e}^{\mu} = g^{\mu \nu} \vec{e}_{\nu}$, where $g^{\mu \nu}=(g^{-1})_{\mu \nu}$ is the inverse of the metric tensor. (We have applied the Einstein summation notation, a sum over any repeated index, one uper and one lower, implicitly.) The choice of dual basis will give the orthonormal property: $\vec{e}^{\mu} \cdot \vec{e}_{\nu} = \delta_{\mu \nu}$.

The change rate of $\vec{e}_\gamma$ in the direction $\beta$ is the vector $\partial_{\beta} \vec{e}_\gamma$ and its projection onto the tangent plane, for the study of the intrinsic geometry, is the covariant derivative 
\begin{equation}
    \nabla_{\beta} \vec{e}_\gamma = {\Gamma^\alpha}_{\beta \gamma} \vec{e}_{\alpha}.
\end{equation}
The coefficients ${\Gamma^\alpha}_{\beta \gamma}$ are called the Christoffel symbol of the second kind. The Christoffel symbols are symmetric in lower indices, ${\Gamma^\alpha}_{\beta \gamma}={\Gamma^\alpha}_{\gamma \beta}$, and the connection is said to be torsion-free. We may need the Christoffel symbol of the first kind 
\begin{equation}
    \Gamma_{\alpha \beta \gamma} = \vec{e}_{\alpha} \cdot \partial_{\beta} \vec{e}_{\gamma} = \trace \left[ (\partial_{\alpha}\hat{P}) (\partial_{\beta}\partial_{\gamma}\hat{P}) \right]; 
    \label{christoffel}
\end{equation}
they are related through the metric tensor:
\begin{equation}
    {\Gamma^\alpha}_{\beta \gamma} = g^{\alpha \kappa}  \Gamma_{\kappa \beta \gamma}.
\end{equation}
The Christoffel symbols can be derived from the derivatives of the metric tensor: 
\begin{equation}
\Gamma_{\alpha \beta \gamma} = \frac{1}{2} \left( g_{\alpha \beta, \gamma} + g_{\alpha \gamma, \beta} - g_{\beta \gamma, \alpha} \right),
\end{equation}
or inversely, 
\begin{equation}
    \partial_\mu g_{\alpha\beta} = \Gamma_{\alpha\beta\mu} + \Gamma_{\beta\alpha\mu}.
\end{equation}
With the connection on the basis frames, the covariant derivative of a vector $\vec{V}=V^\lambda \vec{e}_\lambda$ will be 
\begin{equation}
    \nabla_{\mu} \vec{V} = \left( \partial_\mu V^\lambda +V^{\rho} {\Gamma^{\lambda}}_{\mu \rho} \right) \vec{e}_{\lambda}.
\end{equation}
The Riemann curvature tensor (for torsion-free) involves the second covariant derivatives and is defined as
\begin{equation}
\begin{aligned}
    {R^{\kappa}}_{\lambda \mu \nu} &= \vec{e}^{\kappa} \cdot \left( \nabla_{\mu}\nabla_{\nu} - \nabla_{\nu}\nabla_{\mu}\right) \vec{e}_{\lambda} \\
    &= \partial_\mu {{\Gamma^\kappa}}_{\lambda \nu} - \partial_\nu {\Gamma^\kappa}_{\lambda \mu} + {\Gamma^\kappa}_{\sigma \mu} {\Gamma^\sigma}_{\lambda \nu} - {\Gamma^\kappa}_{\sigma \nu} {\Gamma^\sigma}_{\lambda \mu}, \\
    &= g^{\kappa \rho} \left[ \partial_\mu \Gamma_{\rho \lambda \nu} - \partial_\nu \Gamma_{\rho \lambda \mu} + {\Gamma^\sigma}_{\lambda \mu} \Gamma_{\sigma \rho \nu} -  {\Gamma^\sigma}_{\lambda \nu} \Gamma_{\sigma \rho \mu} \right]
\end{aligned}
\end{equation}
In the last equation, we use
\begin{equation}
\begin{aligned}
    \partial_\mu {\Gamma^\kappa}_{\lambda \nu} &= \partial_\mu \left( g^{\kappa \rho} \Gamma_{\rho \lambda \nu} \right) \\
    &= g^{\kappa \rho} \left( \partial_\mu \Gamma_{\rho \lambda \nu} \right) + \left( \partial_\mu g^{\kappa \rho} \right) \Gamma_{\rho \lambda \nu} \\
    &= g^{\kappa \rho} \partial_\mu \Gamma_{\rho \lambda \nu} -  g^{\kappa \rho} \left(\partial_\mu g_{\rho \delta } \right) g^{\delta \sigma}  \Gamma_{\sigma \lambda \nu} \\
    &= g^{\kappa \rho} \partial_\mu \Gamma_{\rho \lambda \nu} - g^{\kappa \rho} g^{\delta \sigma} \left( \Gamma_{\rho \delta \mu} + \Gamma_{\delta \rho \mu} \right) \Gamma_{\sigma \lambda \nu} \\
    &= g^{\kappa \rho} \left( \partial_\mu \Gamma_{\rho \lambda \nu} -{\Gamma^\sigma}_{\rho \mu} \Gamma_{\sigma\lambda \nu}\right)- {\Gamma^\kappa}_{\delta \mu} \Gamma^\delta_{\lambda \nu}  
\end{aligned}
\end{equation}

The lowered-indexed Riemann curvature is $ R_{\rho \lambda\mu\nu} = g_{\rho \kappa } {R^{\kappa}}_{ \lambda \mu \nu}$, giving 
\begin{equation}
    R_{\rho \lambda\mu\nu} = \partial_{\mu} \Gamma_{\rho\lambda\nu} - \partial_{\nu} \Gamma_{\rho\lambda\mu} + {\Gamma^\sigma}_{\lambda \mu} \Gamma_{\sigma \rho \nu} -  {\Gamma^\sigma}_{\lambda \nu} \Gamma_{\sigma \rho \mu}.
    \label{riemann}
\end{equation}
The expression is not straightforward to understand. When writing them in terms of basis vectors, we have
\begin{equation}
    \partial_{\mu} \Gamma_{\rho\lambda\nu} - \partial_{\nu} \Gamma_{\rho\lambda\mu} = \partial_{\mu} \vec{e}_\rho \cdot \partial_{\nu} \vec{e}_\lambda -  \partial_{\nu} \vec{e}_\rho \cdot \partial_{\mu} \vec{e}_\lambda ,
    \label{Riemann_1}
\end{equation}
and
\begin{equation}
    {\Gamma^\sigma}_{\lambda \mu} \Gamma_{\sigma \rho \nu} -  {\Gamma^\sigma}_{\lambda \nu} \Gamma_{\sigma \rho \mu}=\left( \partial_\mu \vec{e}_{\lambda} \cdot \vec{e}^\sigma \right)
    \left( \vec{e}_\sigma \cdot \partial_\nu \vec{e}_{\rho} \right) - ( \mu \leftrightarrow \nu).
\end{equation}
We find that these two terms are similar but do not cancel with each other because the frame $\{\vec{e}_{\sigma}\}_{\sigma=1}^k$ is just complete ($k=2$ for most of time) for the tangent plane but not for the ambient Euclidean space. We assume the normal basis set that is perpendicular to the tangent plane to be $\{\vec{n}_{\sigma}\}_{\sigma=1}^{N^2-k}$. The completeness relation is written as
\begin{equation}
\mathbb{I} = \sum_{\sigma=1}^k\vec{e}^{\sigma} \otimes \vec{e}_{\sigma} + \sum_{\sigma'=1}^{N^2-k}\vec{n}^{\sigma'} \otimes \vec{n}_{\sigma'}.
\label{completeness}
\end{equation}
With this invention, the Riemann curvature becomes
\begin{equation}
R_{\rho \lambda\mu\nu} =\left(   \partial_\mu \vec{e}_{\rho} \cdot \vec{n}^{\sigma'} \right)
    \left( \vec{n}_{\sigma' }\cdot \partial_\nu \vec{e}_{\lambda} \right) - ( \mu \leftrightarrow \nu).
    \label{Riemann_curv}
\end{equation}
(Keep in mind that $\sigma'$ runs over degrees in the normal plane.) There are symmetries in the Riemann curvature tensor that restrict the number of independent components. In two dimensions of $\vb{k}$, the only independent component is
\begin{equation}
    R_{1212} = \sum_{{ij}=\{x,y\}} \epsilon_{ij} \left(\partial_x \vec{e}_{i} \cdot \vec{n}^{\sigma'}\right)
    \left( \vec{n}_{\sigma' }\cdot \partial_y \vec{e}_{j} \right),
    \label{R1212}
\end{equation}
where $\epsilon_{ij}$ is the Levi-Civita symbol. 
This formula is very similar to the Berry curvature, which is
\begin{equation}
\begin{aligned}
    \Omega_{\alpha} &= \Im \sum_{{ij}=\{x,y\}} \epsilon_{ij} \matrixel{\partial_i u_{\alpha}}{\hat{Q}_\alpha}{\partial_j u_{\alpha}}  \\
    &= \Im \sum_{{ij}=\{x,y\}} \epsilon_{ij} \trace \left[\left( \partial_i\hat{P}_\alpha\right) \hat{Q}_\alpha \left( \partial_j\hat{P}_\alpha\right)  \right],
\end{aligned}
\end{equation}
where $\hat{Q}_\alpha = 1 - \hat{P}_\alpha$. The comparison tells that the former one concerns the change in the tangent vectors, while the latter concerns the coordinates; mathematically, the former involves second-order derivatives, while the latter involves first-order ones. 

The Riemann curvature is related to the Gauss curvature $K_G$ by
\begin{equation}
    K_G = \frac{R_{1212}}{\det g}.
\end{equation}
The Gauss-Bonnet theorem, which is 
\begin{equation}
    \int_{\widetilde{M}} K_G \, dA = 2\pi \chi(\widetilde{M}),
    \label{GB1}
\end{equation}
states that the integral of the Gauss curvature over a closed surface $\widetilde{M}$ reflects the topology of the surface. Here $\widetilde{M}$ will be the manifold of the quantum states, $dA= \abs{\vec{e}_1 \times \vec{e}_2} \, dk_x \wedge dk_y= \sqrt{\det g} \, dk_x \wedge dk_y$ is the element area form, and $\chi(\widetilde{M})$ is the Euler characteristic of $\widetilde{M}$. The formula is reminiscent of the Chern number $C_\alpha \in \mathbb{Z}$ as the integral of the Berry curvature:  
\begin{equation}
    \int_{M} \Omega_{\alpha} \, dk_x dk_y = 2\pi C_\alpha. \label{int_Berry}
\end{equation}
Here $M$ is Brillouin zone (BZ). The relation between these two integrals will be discussed later.

In this work, we study a two-band model with the Bloch Hamiltonian
\begin{equation}
    H_{\mathrm{2band}}(\vb{k}) = \vec{d}(\vb{k})  \cdot \vec{\sigma},
\end{equation}
where $\sigma$'s are Pauli matrices and 
\begin{equation}
\begin{aligned}
    d_x &=  \sin k_x,\\
    d_y &=  \sin k_y, \\
    d_z &= m_0 -\cos k_x -\cos k_y
\end{aligned} \label{eq:d_vec}
\end{equation}
with $m_0=1$ for a topological insulator of the Chern number $C=1$. 
The eigenprojector of the lower-band is $\hat{P}=\frac{1}{2}\left( \mathbb{I} -\vec{\sigma}\cdot \hat{d} \right)$, where $\hat{d}=\vec{d}/d$ ($d=|\vec{d}|$) is a unit vector. 
This describes the map $f: M \rightarrow \widetilde{M} \; (\vb{k} \mapsto \hat{P})$, from a torus $M=T^2$ to a sphere $\widetilde{M}=S^2$.  

\begin{figure}[tb]
\centering
\includegraphics[width=0.4\textwidth]{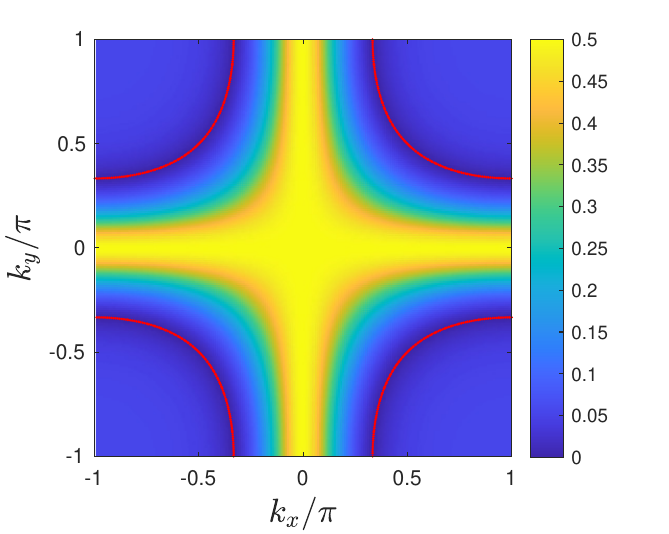}
\caption{The determinant of the metric tensor $\det g$ for the two-band model. The red lines are singular points where $\det g=0$.}
\label{fig:detg}
\end{figure}

For our model $H_{\mathrm{2band}}$, we choose the normal vectors as $\vec{n}_1=\vec{n}^1 = \hat{P}$ and $\vec{n}_2=\vec{n}^2 = \hat{1}-\hat{P}$, which correspond to the eigenprojectors of the valence and conduction bands, respectively. This choice is justified by the idempotent property of $\hat{P}$, which ensures that $\trace (\hat{P} \partial_\mu \hat{P})=0$. 
A direct calculation using Eq. (\ref{R1212}) yields that $R_{1212} =2\det g$. Consequently, the Gauss curvature is constant, $K_G=2$, indicating that the manifold is a sphere of radius $1/2$. However, the system exhibits complexities beyond our expectation. As shown in Fig.~\ref{fig:detg}, the function $\det g$ contains singular points--highlighted along the red line--where $\det g =0$. At these singularities, the Christoffel symbols of the second kind, and hence the Riemann curvature tensor, become ill-defined due to the requirement of $g^{-1}$. Therefore, the Gauss-Bonett formula in Eq. (\ref{GB1}) may not be applicable to this model. Even if we exclude the singular points, integrating the constant curvature $K_G=2$ does not yield a quantized result, as the area element $\sqrt{\det g}$ is not uniform across the manifold. A theoretical framework to address these singularities is essential and will be developed in the following section.

\section{singular points}
\subsection{singular curve} \label{subsec:singular_curve}
\begin{figure}[tb]
\centering
\includegraphics[width=0.45\textwidth]{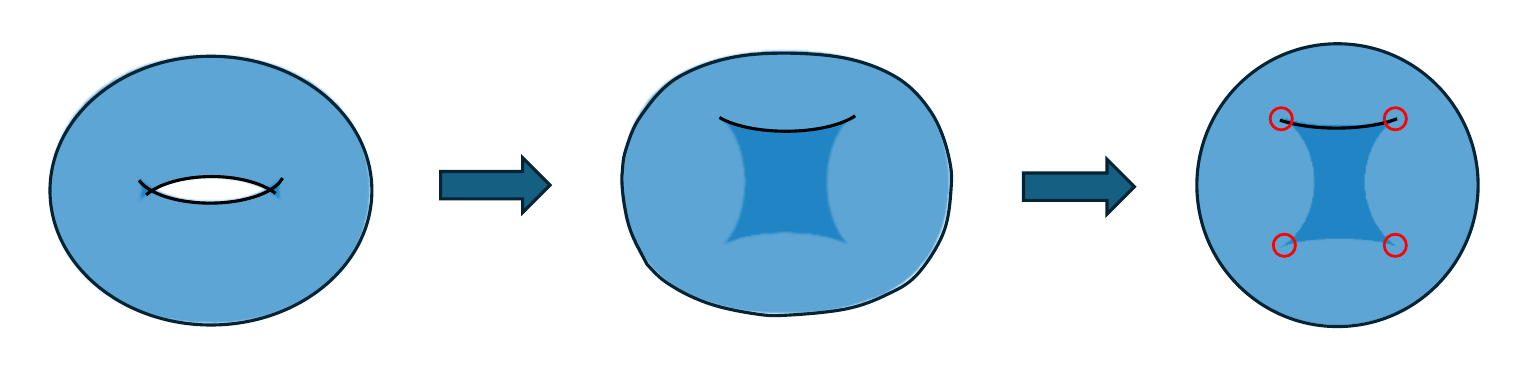}
\caption{A transformation from a torus to a sphere. In the transparent tori, the dark areas are the front face made of three projections. In the most right object, the Whitney cusps are marked by red circles. }
\label{fig:whitney}
\end{figure}

It has been shown that for a two-band model, singular points of the metric tensor must exist~\cite{ozawa2021,mera2021,wang2021}. This arises because a mapping from a torus (the momentum manifold) to a sphere (the quantum- state manifold) is not an immersion~\cite{mera2021}, meaning the differential map is not injective. Figure \ref{fig:whitney} illustrates a simple morphism from a torus to a sphere, clearly showing the unavoidable intersections and Whitney cusps \cite{Arnold1988}.

The singular or critical point is defined as a point where the metric tensor is degenerate, i.e., $\det (g_{\mu \nu}) =0$. This is equivalent to one of the metric tnesor's eigenvalues being zero. At a singular point, the rank of the map is lower than the momentum dimension~\cite{arnold1985}. In the specific case of rank-1, a kernel vector field $\eta$ is defined such that $\partial_{\eta}\hat{P}_\alpha=\eta^{\mu}\partial_{\mu}\hat{P}_\alpha=0$. The nonzero solution $(\eta^1,\eta^2)^\mathsf{T}$ is the kernel eigenvector of $g_{\mu \nu}$. 

\begin{figure}[tb]
\centering
\includegraphics[trim=18 0 18 0, clip, width=0.5\textwidth]{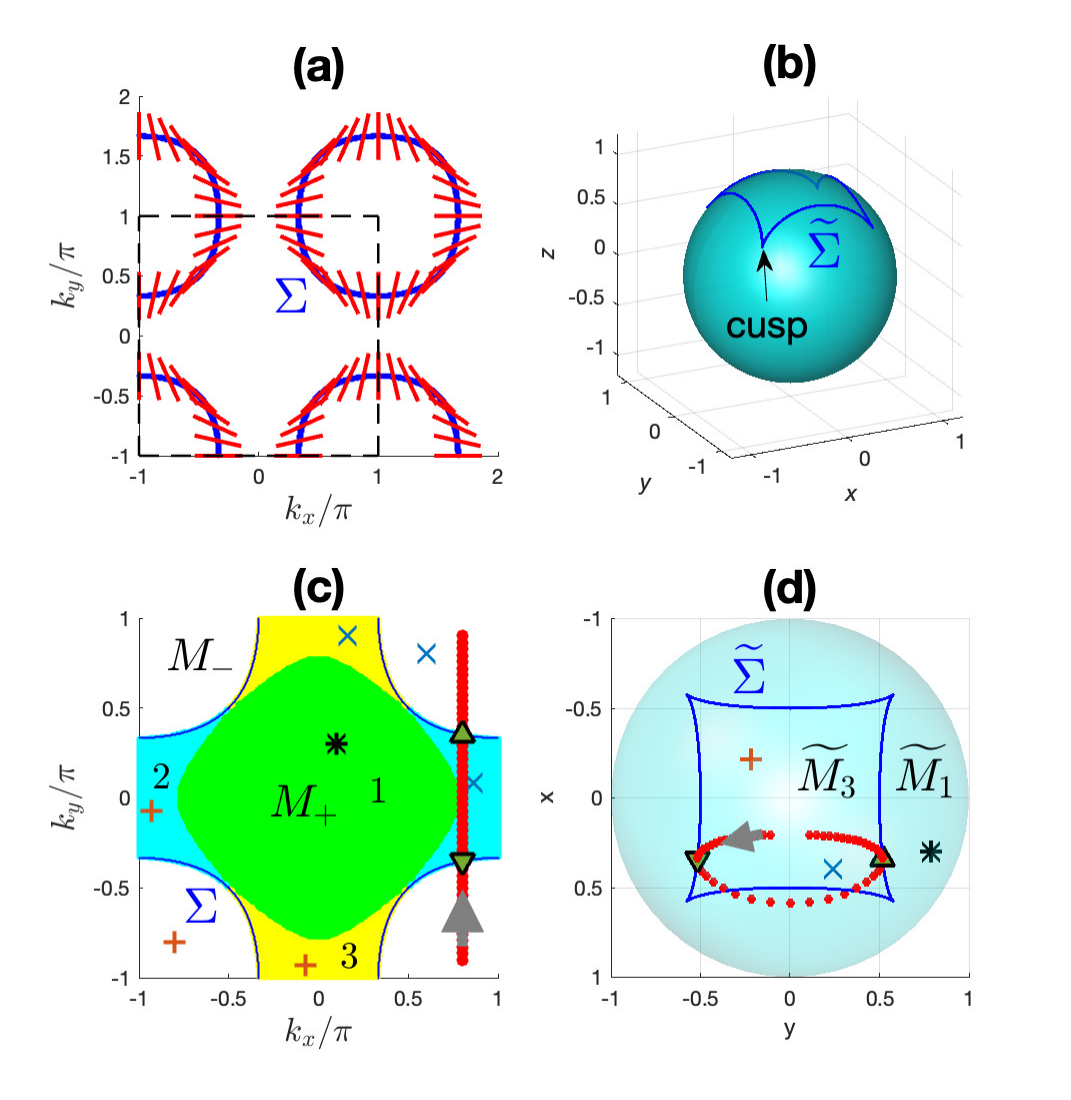}
\caption{(a) The singular curve (blue lines) and the kernel vectors (in red) at singular points. The dashed lines circumscribe the first Brillouin zone. (b) The manifold of the quantum system (the sphere) and the singular-value curve (the blue line). Four cusp points are visible as the image of $(k_x,k_y)=\pm(\pi/2,\pm\pi/2)$. (c) and (d) are the $\vb{k}$ points and their maps, respectively. The domain is separated into $M_+$ and $M_-$. In addition, $M_+$ is the union of three colorful regions (1, 2, and 3). A trajectory in $k$-space (red lines) in the direction as the gray arrow shows is in (c) and its image on the sphere (red dots) is in (d). When the curve passes the singular points (green triangles), the image curve is deflected at $\widetilde{\Sigma}$. Each point in $\widetilde{M}_3$ has three preimages (illustrated by blue $\times$ and red $+$ markers), while every point in $\widetilde{M}_1$ has one preimage (by black $\ast$). }
\label{fig:sing_pts}
\end{figure}

The set of the singular points, denoted as $\Sigma$, forms a continuous closed curve that encloses $(\pi,\pi)$. The curve divides $M$ into two regions, $M_+$ and $M_-$, as shown in Fig.~\ref{fig:sing_pts}(c). We can express the BZ as $M=M_+\sqcup M_-\sqcup \Sigma$. As illustrated by the red lines in Fig.~\ref{fig:sing_pts}(a), the kernel field completes one full rotation along this curve. At the peak points $(k_x,k_y) = \pm (\pi/2,\pi/2)$ and $\pm (\pi/2,-\pi/2)$, the kernel field is tangent to the singular curve, yielding $\eta_1=-\eta_2$ and $\eta_1=\eta_2$, respectively. 
The image of a singular point is called a singular value. The set of these values, $\widetilde{\Sigma}=f(\Sigma)$, forms a closed contour too. This contour, represented by the blue lines in Figs.~\ref{fig:sing_pts}(b), (d), divides the $S^2$ manifold into two open regions, $\widetilde{M}_1$ and $\widetilde{M}_3$. We denote the image manifold as $\widetilde{M}=\widetilde{M}_1 \sqcup \widetilde{M}_3 \sqcup \widetilde{\Sigma}$.
The singular values at the peak points are cusp points (swallowtails), where the external angle of the image tangent changes by, $\abs{\theta} = \pi$. The cusp number is related to the topologies of the domain and the image~\cite{izumiya1993}.

The singular points are loci of folds. Figure~\ref{fig:sing_pts}(c) shows a curve (in red) that traverses vertically and crosses two singular points (marked by triangles). The corresponding image of the curve deflects or ``folds back" at the singular values as shown in Fig.~\ref{fig:sing_pts}(d). This folding implies a many-to-one mapping. Specifically, each point in $\widetilde{M}_1$ has a single preimage, while each point in $\widetilde{M}_3$ has three preimages. This can be visualized in Fig. \ref{fig:sing_pts}(d), where a point marked with an asterisk ($\ast$) in $\widetilde{M}_1$ and two points marked with a plus sign ($+$) and an cross ($\cross$) in $\widetilde{M}_3$ are shown. Their corresponding preimages, with the same symbols, are shown in Fig. \ref{fig:sing_pts}(c).
To simplify the description of the preimages, the set $M_+$ can be considered as a combination of three subsets: $M_{+,1}$, $M_{+,2}$, and $M_{+,3}$. The latter two are open sets, whereas $M_{+,1}$ contains the boundaries to both $M_{+,2}$ and $M_{+,3}$.
The maps of $M_{+,2}$ and $M_{+,3}$ both project into $\widetilde{M}_3$. In contrast, the map of $M_{+,1}$ projects into the region $\widetilde{M}_1 \cup \tilde{\Sigma}-\mathrm{(4 ~cusps)}$, where the image of the boundary of $M_{+,1}$ (to $M_{+,2}$ and $M_{+,3}$) is $\widetilde{\Sigma}-\mathrm{(4 ~cusps)}$.

\subsection{signed area form}
For a regular surface, the orientation (or handedness) of the coordinate system is fixed across the entire surface. However, in the present of singularities, the map's orientation changes sign on either side of the singular curve $\Sigma$. We understand the orientation from the differential map. If a positive frame $(k_x,k_y)$ is chosen in the domain $M$, the corresponding basis vectors of the differential map, $(\vec{e}_1,\vec{e}_2)$, do not maintain the same handedness over the entire surface. This is a direct consequence of the metric tensor becoming degenerate at the singular points. The relationship $\abs{\vec{e}_1 \cross \vec{e}_2}^2 = \det g$ shows that when $\det g$ approaches zero, the cross product of the basis vectors also approaches zero, which implies that the angle between $\vec{e}_1$ and $\vec{e}_2$ becomes zero. Since the map is continuous, it is expected that the handedness of the frame changes as it crosses a critical point.

To address the problem of singularities, the concept of a front is used. A map $f$ from a two-dimensional domain to $\mathbb{R}^3$ is defined as a \emph{front} if there exists a unit normal vector field, $\vec{N}$, that is orthogonal to the tangent vector field of the map at every point in the domain. In this sense, $\vec{N}$ can be considered a Gauss map of $f$. 
The key advantage of a front is that the pair $(f,\vec{N})$ constitutes an immersion, even when nondegenerate singularities exist. The presence of the normal vector allows for the definition of a limiting tangent plane and its corresponding basis.
For our two-band model, the map is a front because a normal vector is readily found to be $\pm \hat{P}$, which is orthogonal to the tangents due to the idempotency of the projector $\hat{P}$. This property ensures that the map satisfies the definition of a front, providing a framework for analyzing its singular behavior.

In this extrinsic viewpoint, the orientation is defined by $\mathrm{sgn}\left(\vec{N}\cdot \vec{e}_1 \cross \vec{e}_2 \right)$, where $\vec{N}$ is the surface's unit normal vector. Considering the orientation factor, the \emph{signed area form} is defined as 
\begin{equation}
    d\bar{A}= \bar{\lambda} \, dk_x \wedge dk_y,
\end{equation}
where the \emph{signed area density} in the eigenprojector formalism is
\begin{equation}
    \bar{\lambda} = -i \trace \left( \hat{P} \,[\partial_x\hat{P},\partial_y\hat{P}] \right).
    \label{eq:signed_density}
\end{equation}
The imaginary number is included for Hermiticity. This choice gives $\bar{\lambda}>0$ in $M_+$ and $\bar{\lambda}<0$ in $M_-$. It can be verified that $\abs{\bar{\lambda}} = \sqrt{\det g}$. The signed area density is preferable to the standard area density $\sqrt{\det g}$ because it remains differentiable even at singular points.

\subsection{singular curvature}
In the next subsection, we introduce the Gauss–Bonnet theorem for manifolds with boundary. At the boundary, it is necessary to measure the geodesic curvature. Along the singular curve, the adapted coordinates $u=u(k_x,k_y)$ and $v=v(k_x,k_y)$ prove useful. The corresponding directional derivatives are $\partial_u = \frac{\partial k^\mu}{\partial u} \partial_{\mu}:=u^\mu \partial_{\mu}$ and $\partial_v = \frac{\partial k^\mu}{\partial v} \partial_{\mu}:=v^\mu \partial_{\mu}$.
We designate $\vec{u}$ as the tangent vector field along the singular curve (oriented clockwise), and $\vec{v}$ as a kernel field, forming a positively oriented frame $(u,v)$ (see Fig.~\ref{fig:uv_norm}). The solutions for the singular curve and the adapted coordinates are provided in Appendix~\ref{appendix:apated_coor}.

The induced tangent vectors are defined: $\vec{e}_u  = u^{\mu} \vec{e}_{\mu} $ and $\vec{e}_v \dot  = v^{\mu} \vec{e}_{\mu}$. (Note that this is equivalent to, for example, $\partial_{u}\hat{P}=u^{\mu}\partial_{\mu}\hat{P}$.) 
The dual vectors are $\vec{e}^{\,u} = u_{\mu} \vec{e}^{\,\mu}$ and $\vec{e}^{\,v} = v_{\mu} \vec{e}^{\,\mu}$. When $u$ and $v$ are linearly independent, $u\wedge v=u^1v^2 - u^2 v^1 \neq 0$, 
\[
\mqty(u_1 & u_2 \\ v_1 & v_2) = \frac{1}{u\wedge v} \mqty(v^2 & -v^1 \\ -u^2 & u^1), 
\]
which gives $\vec{e}_u \cdot \vec{e}^u = \vec{e}_v \cdot \vec{e}^v = 1$ and $\vec{e}_u \cdot \vec{e}^v = \vec{e}_v \cdot \vec{e}^u = 0$ based on $\vec{e}_{\mu} \cdot \vec{e}^{\nu} = \delta_{\mu}^{\nu}$. The completeness implies $\sum_{\mu} \vec{e}^{\mu} \otimes \vec{e}_{\mu} =\vec{e}^{u} \otimes \vec{e}_{u} + \vec{e}^{v} \otimes \vec{e}_{v}$.
For convenience, we define the metric components as $E=\vec{e}_u \cdot \vec{e}_u$, $G=\vec{e}_v \cdot \vec{e}_v$, and $F=\vec{e}_u \cdot \vec{e}_v$.

Curvature can be interpreted as the acceleration of a curve, measured with respect to its arc length parameter. Along the singular curve, the unit tangent vector is given by $\gamma'=\frac{1}{\sqrt{E}}\vec{e}_u$. 
The covariant derivative of the tangent vector in the direction of $u$ is computed as
\begin{equation}
\begin{aligned}
    \gamma''&=\frac{1}{\sqrt{E}} \nabla_u \gamma' \\
    &=  
    \frac{1}{\sqrt{E}} \left( \partial_u \frac{1}{\sqrt{E}}\right)\vec{e}_u + \frac{1}{E}\left(\vec{e}^\rho\cdot \partial_{u}\vec{e}_u \right) \vec{e}_\rho \\    
    &= \frac{1}{E} \left[ \left(-\frac{1}{2} \partial_{u} \log E + {\Gamma^u}_{u u} \right) \vec{e}_u + {\Gamma^v}_{u u} \vec{e}_v \right].  
    \label{gamma_2nd_der}
\end{aligned}
\end{equation}
The choice of the unit vector ensures that $\gamma''$ is orthogonal to $\gamma'$. 
In the final expression, the Christoffel symbols in the $(u,v)$ coordinate basis. The transformation rule for ${\Gamma^v}_{u u}$ is given by
\begin{equation}
    {\Gamma^v}_{u u} = \frac{\partial v}{\partial x^\rho}  \frac{\partial x^\mu }{\partial u} \frac{\partial x^\nu }{\partial u} {\Gamma^\rho}_{\mu \nu} 
    + \frac{\partial v}{\partial x^\rho} \frac{\partial^2 x^\rho }{\partial u^2}.  
\end{equation}

The geodesic curvature along $u$ is the acceleration projected onto the tangent plane and is given by~\cite{do2016differential}
\begin{equation}
    \kappa_\mathrm{g} = \gamma'' \cdot \left( \vec{N} \times \gamma'\right).
\end{equation}
In this expression, the component of $\gamma''$ along $\vec{e}_u$ in Eq. (\ref{gamma_2nd_der}) does not contribute, so the geodesic curvature simplifies to:
\begin{equation}
    \kappa_\mathrm{g} = \frac{1}{E^{3/2}} {\Gamma^v}_{u u} \vec{N} \cdot \left( \vec{e}_u \times \vec{e}_v \right).
\end{equation}
The term $\vec{N} \cdot \left( \vec{e}_u \times \vec{e}_v \right)=\bar{\lambda}(\vec{N},u,v)$ is the signed area density in the adapted coordinate frame, as discussed in Eq. (\ref{eq:signed_density}). Note that the signed area form, $d \bar{A} = \bar{\lambda}(\vec{N},u,v) \, d u\wedge dv $, is invariant under orientation-preserving coordinate transformations. The Christoffel symbol is given by~\cite{do2016differential}:
\begin{equation}
{\Gamma^v}_{u u} = \frac{2EF_u-FE_u-EE_v}{2(EG-F^2)}, 
\end{equation}
where subscripts $u$ and $v$ denote partial derivatives, e.g. $F_u = \partial F/\partial u$. 
This expression becomes indeterminate (of the form $ \frac{0}{0}$) as the curve approaches the singular curve ($v=0$). However, the geodesic curvature remains well-defined in the limit due to the presence of the signed area density $ \bar{\lambda}(\vec{N},u,v)$ in the numerator.
The limiting geodesic curvature will be named the singular curvature $\kappa_s$~\cite{saji2009}:
\begin{equation}
    \begin{aligned}
    \kappa_s &= \frac{2EF_u-FE_u-EE_v}{2 E^{3/2} \bar{\lambda}} \\ 
    &= \frac{2EF_{uv}-F_v E_u-EE_{vv}}{2 E^{3/2} \bar{\lambda}_v}  ,
    \end{aligned}
\end{equation}
where the second line follows from applying L'Hôpital's rule. The derivative of the signed area density is given by:
\begin{equation}
    \bar{\lambda}_v = \frac{\partial \bar{\lambda}(\vec{N},u,v)}{\partial v} 
    = -i \trace \left( \hat{P} \,[\partial_u\hat{P},\partial_v^2\hat{P}] \right).
\end{equation}

To compute the line integral of the singular curvature along the singular curve, the length element is defined to be the differential arc length, $ds = \sqrt{E}du$. In numerical implementations, it may be necessary to express $du$ in terms of $dk_x$ and $dk_y$, which represent the spacings between adjacent $k$-points along the curve. We use the property that a one-form maps a vector to a scalar, i.e., $du(\partial_u)=dk_x(\partial_x)=dk_y(\partial_y)=1$. By expressing $du$ as a linear combination of $dk_x$ and $dk_y$, and using $\partial_u = u^\mu \partial_\mu$, we obtain $du=\frac{1}{u\wedge v}(v^2 dk_x + v^1 dk_y) $.

\section{Gauss-Bonnet theorem}
Since there are singular points where the Gauss curvature is ill-defined, we study the Gauss-Bonnet theorem in two submanifolds, $\widetilde{M}_{\pm} = f(M_{\pm})$, with the singular values as their boundaries. For a manifold with boundary, the theorem is given by:
\begin{equation}
    \int_{\widetilde{M}_{\pm}} K_G \, dA +  \int_{\partial \widetilde{M}_{\pm}} \kappa_\mathrm{g} ds + \sum_{i=1}^4 \theta_i = 2\pi \chi\left(\widetilde{M}_{\pm} \right),
\end{equation}
where $K_G$ is the Gauss curvature, $\kappa_\mathrm{g}$ is the geodesic curvature on the singular curve, $\theta$'s are external angles at the cusps, and $\chi$ is the Euler characteristic. 
We begin by emphasizing that the Gauss–Bonnet theorem is independent of the orientation of the manifold. The Gauss curvature is expressible solely in terms of the metric tensor and thus remains unaffected by the choice of normal vector. When the normal vector of the manifold is reversed, the geodesic curvature changes sign; however, the direction of traversal along the boundary also reverses. These two effects compensate in the line integral of the geodesic curvature, rendering it invariant under orientation changes. Similarly, the external angle depends on both the orientation of the manifold and the direction of the boundary curve~\cite{do2016differential}. Since the boundary curve inherits its direction from the manifold’s orientation, these dependencies compensate each other. Notably, the Euler characteristic is a topological invariant that does not require any information about the manifold’s orientation and is well-defined even for non-orientable manifolds.

With our chosen frame, the curve $\Sigma$ is traversed both clockwise in $\partial M_+$ and $\partial M_-$, and the same is true for its image. [Imaging that the normal vector is pointing outward, the boundary is traversed so that the interior of $\widetilde{M}_{\pm} = f(M_{\pm})$ is on the left (right) of the boundary. The handedness is reflected in the defintion of the geodesic curvature through $\vec{N} \times \gamma'$.] The geodesic curvatures as well as their line integrals on $\widetilde{M}_{\pm}$ will converge to the singular curvature when their corresponding curves approach the singular curve. Therefore, the sum and difference of the integrals yield:
\begin{equation}
    \int_{\partial\widetilde{M}_{+}} \kappa_\mathrm{g} ds \pm \int_{\partial \widetilde{M}_{-}} \kappa_\mathrm{g} ds = 
    \begin{cases}
    2 \int_{\widetilde{\Sigma}} \kappa_\mathrm{s} ds \\
        0                                  
    \end{cases} \: .
\end{equation}

As for the external angle, it follows the right-hand (left-hand) rule in $\widetilde{M}_+$ ($\widetilde{M}_-$). Referring to Fig.~\ref{fig:sing_pts}, for a clockwise trajectory the external angle at each cusp in $\widetilde{M}_+$ is $-\pi$, while it is $+\pi$ in $\widetilde{M}_+$.

Based on the above analysis, summing and subtracting the integrals for the two submanifolds gives two distinct formulas:
\begin{equation}
    \frac{1}{2\pi} \left(\int_{\widetilde{M}} K_G \, dA + 2\int_{\widetilde{\Sigma}} \kappa_s ds \right) = \chi\left(\widetilde{M}_+\right)+  \chi\left(\widetilde{M}_-\right),
    \label{eq:sum1}
\end{equation}
and
\begin{equation}
    \frac{1}{2\pi}\int_{\widetilde{M}} K_G \, d\bar{A} = \chi\left(\widetilde{M}_+\right)-  \chi\left(\widetilde{M}_-\right) + 4.
    \label{eq:sub1}
\end{equation}
It is seen that the cusps have no effect in the sum term while they give 4 in the subtraction term. The right-hand sides involve the Euler characteristics.
We evaluate the Euler characteristic for each map. As discussed previously in Sec. \ref{subsec:singular_curve}, the Euler characteristic for a disjoint union is the sum of the characteristics of its components. Thus, $\chi(\widetilde{M}_+)  = \chi \left[ S^2 - \mathrm{(4 ~cusps)}) \right] + \chi \left(\widetilde{M}_3\right)$ and $\chi(\widetilde{M}_-)=\chi(\widetilde{M}_3)$. 
The punctured sphere has the Euler characteristic $\chi \left[ S^2 - \mathrm{(4 ~cusps}) \right]=2-4=-2$, and $\widetilde{M}_3$, being topologically equivalent to a disk, has $\chi(\widetilde{M}_3)=1$, so the sum and the difference are 
\[ 
\chi\left(\widetilde{M}_+\right)+  \chi\left(\widetilde{M}_-\right)=0 \]
and 
\[ 
\chi\left(\widetilde{M}_+\right)-  \chi\left(\widetilde{M}_-\right)=-2.
\]
Therefore, the right-hand sides of Eqs.~(\ref{eq:sum1}) and Eqs.~(\ref{eq:sub1}) evaluate to $0$ and $2$, respectively. 

These formulas are consistent with our numerical calculations. 
Our calculated values for the integral of the Gauss curvature is $2.3779\times 2\pi$, which indicates that the total area, quantum volume,  is half the value: $\int_{\widetilde{M}} dA=1.1889 \times 2\pi$. The qunatum volume is larger than $2\pi$, which is the area of a sphere of radius $1/K_G=0.5$. The excess area is due to the folding. The area in $\widetilde{M}_3$ is found to be $\int_{\widetilde{M}_3} dA = 0.0945 \times 2\pi$. By subtracting twice the area in $\widetilde{M}_3$ from the total area, we obtain a quantized value of $2\pi$. Including $K_G=2$, this result is consistent with the right-hand side of Eq. (\ref{eq:sub1}). 

\begin{figure}[tb]
\centering
\includegraphics[width=0.45\textwidth]{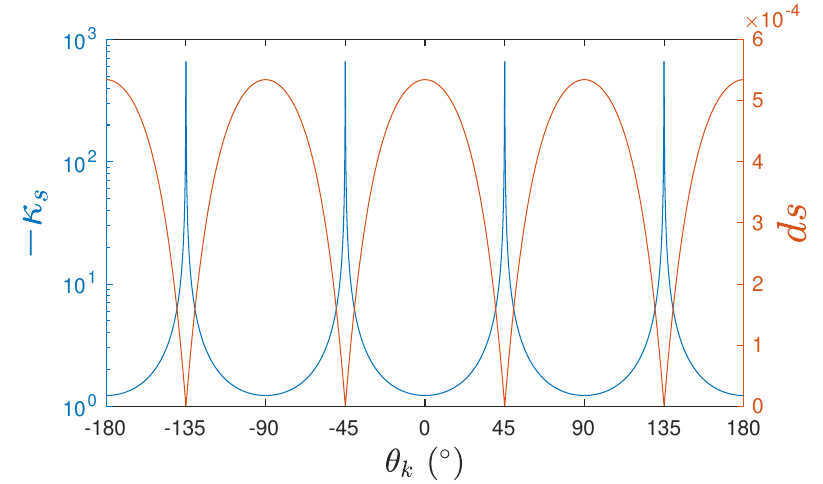}
\caption{The singular curvature $\kappa_s$ and the length element $ds$ in the line integral along the singular curve. The angle $\theta_k$ is measured relative to the $k_x$-axis around $(\pi,\pi)$. While the singular curvature diverges at the peaks, the length element simultaneously approaches zero.}
\label{fig:singular_curvature}
\end{figure}

For Eq. (\ref{eq:sum1}), our numerical result shows that the integral of the doubled singular curvature almost exactly cancels out the integral of the Gaussian curvature, consistent to the topological argument. Our obtained values are $-2.3710$, $-2.3745$, and $-2.3756$ for 400, 800, and 1200 grid points on the singular curve, respectively. The integral value is sensitive to the number of grid points because the singular curvature diverges at the cusps. The singular curvature and the length element along the singular curve are shown in Fig.~\ref{fig:singular_curvature}. The divergence behavior and the negative sign of the singular curvature agree with the claims made in Ref. \cite{saji2009}.        

Finally, we note that these theorems can be generalized for a general Gauss map with a degree of $\deg(f)=C_\alpha$. Given that the existence of Whitney cusps implies a triple covering, it is suggested that $\chi(\widetilde{M}_+)  = C_\alpha \chi(S^2 \sqcup \mathrm{disk}) - 4$. The $-4$ correction arises from the removal of the four cusp points. Substituting these into the formulas yields:
\begin{equation}
    \frac{1}{2\pi} \left(\int_{\widetilde{M}} K_G \, dA + 2\int_{\widetilde{\Sigma}} \kappa_s ds \right) = 4 (C_\alpha -1),
    \label{eq:sum2}
\end{equation}
and
\begin{equation}
    \frac{1}{4\pi}\int_{\widetilde{M}} K_G \, d\bar{A} = C_\alpha.
    \label{eq:sub2}
\end{equation}
The last formula is a key result of this study and it was proposed in mathematics field \cite{Kossowski2002,saji2008,saji2009}. 

Compared to Eq.~(\ref{int_Berry}), this suggests that the signed area density in Eq. (\ref{eq:signed_density}) is exactly the Berry curvature, giving 
\begin{equation}
    \Omega =\bar{\lambda } = -\frac{1}{2} \hat{d} \cdot 
    \left(\partial_x\hat{d} \cross \partial_y\hat{d}\right),
\end{equation}
and that the Chern number measures the total signed area swept out by the quantum state on the Bloch sphere or projective Hilbert space. 
The Berry curvature in terms of eigenprojectors was proposed in many references~\cite{avron1985,bellissard1994,panati2007,prodan2009,monaco2017,mera2021kahler}. This coincidence leads to the coercive relation between the metric tensor and the Berry curvature in a two-band model: $\det g = \Omega^2$~\cite{mera2021kahler,mera2022nontrivial,wang2021,peotta2015}. 
We remark that the eigenprojector representation of the Berry curvature in Eq. (\ref{eq:signed_density}) suits for any model, but it is applied to the signed area form only for the two-band model. The notion of a front relies on the existence of a unique normal direction at each point. In the quantum-geometry context, the connection between the metric tensor and the Berry curvature is still mysterious. The quantum metric is the upper bound of the Berry curvature is also an interesting phenomenon~\cite{roy2014band,jackson2015,peotta2015,Mitscherling2022,mera2022relating,Herzog2022,kwon2024,yu2025quantum}. Our finding indicates that singular folds appears in the quantum-state manifold, and the folding structure might result in the discrepancy between the quantum volume and the Chern number~\cite{mera2022relating}.

\section{Conclusion}
In summary, we have successfully introduced and rigorously analyzed the concept of the quantum Gauss curvature $K_G$ for two-dimensional Bloch bands, utilizing the interpretation of the eigenprojector $\hat{P}$ as local coordinates of a surface in the projective Hilbert space (the Grassmannian). Building on the formalism of Graf and Piéchon's work \cite{graf2021}, we expressed the eigenprojector as a polynomial of the Bloch Hamiltonian, enabling analytic evaluation of its derivatives. 
By deriving the full Riemann curvature tensor from the quantum metric tensor, we obtained a local, gauge-invariant expression for $K_G$ that quantifies the intrinsic curvature of the quantum state manifold. Our analysis demonstrates that the Gauss curvature provides a valuable local perspective on the geometry of the band structure, complementing the Berry curvature, which is primarily related to global topology. In a two-band model, we identify the emergence of fold and cusp singularities where the metric tensor degenerates. Treating the manifold as a front with a well-defined normal vector field, we defined a singular curvature along the fold and derived a generalized Gauss–Bonnet theorem that remains valid in the presence of these singularities. The resulting relation links the total signed Gauss curvature to the Chern number, revealing that the Berry curvature is the signed area density of the quantum manifold. 
For multiband systems, where the mapping is no longer frontal, the geometric identification becomes subtler and remains an open question. Future work may explore how this geometric invariant manifests in measurable quantities—possibly through higher-order response functions—and how it extends to systems with additional internal symmetries or non-Hermitian band structures. The framework developed here establishes a geometric foundation for understanding the curvature and topology of quantum states and offers new avenues for connecting geometry to experimentally accessible observables.

\section{Acknowledgements}
This work was supported by the National Science and Technology Council (NSTC) under grant number 114-2112-M-110-018. The author also thank the support from NCTS.

\appendix
\section{Derivatives of Eigenprojectors}
\subsection{Single band}
To calculate the derivatives of the eigenstate projectors, we need their analytic function forms. However, most of the time, we cannot have them when the size of the Hamiltonian is large, $N>2$. Fortunately, according to the Cayley-Hamilton theorem, any power or analytic function of a matrix of size $N\times N$ can be expressed as a linear combination of lower powers of that matrix up to the power of $N-1$. It implies that the eigenprojector can be expressed as a polynomial of the Hamiltonian. Graf and Piéchon had derived the formula \cite{graf2021}, and we will borrow their idea and show it here. 

For the $\alpha$-th eigenstate projector, it is written as
\begin{equation}
    \hat{P}_{\alpha} = \prod_{\beta \neq \alpha} \left( \frac{H-E_\beta 1_{N}}{E_{\alpha} - E_{\beta}} \right). 
    \label{Pa}
\end{equation}
The Hamiltonian as a function of $\vb{k}$ is known, but the eigenenergies $E_\beta$ might be obtained numerically.
The numerator of Eq. (\ref{Pa}) is expanded to be
\begin{equation}
\begin{aligned}
   \hat{K}_{\alpha} &=\prod_{\beta \neq \alpha}  \left( H-E_\beta 1_{N} \right) \\
    &=\sum_{k=0}^{N-1} (-1)^k e_k(E_1, \cdots, E_{\alpha-1}, E_{\alpha+1},\cdots,E_N)H^{N-1-k},
\end{aligned}    \label{Ka0}
\end{equation}
where $e_k(E_1, \cdots, E_{\alpha-1}, E_{\alpha+1},\cdots,E_N)$ is the elementary symmetric polynomials of degree $k$ in variables $E_1, \cdots, E_{\alpha-1}, E_{\alpha+1}$. Since the polynomials contain only $N-1$ variables without $E_{\alpha}$, it is better to replace them by ones of all $N$ variables. By observation, we have
\begin{equation}
\begin{aligned}
    e_k(E_1, \cdots, E_N) &= 
    e_k(E_1, \cdots, E_{\alpha-1}, E_{\alpha+1},\cdots,E_N) \\
    &+ E_\alpha e_{k-1}(E_1, \cdots, E_{\alpha-1}, E_{\alpha+1},\cdots,E_N),
\end{aligned}
\end{equation}
which can lead to 
\begin{equation}
\begin{aligned}
    e_k(&E_1, \cdots, E_{\alpha-1}, E_{\alpha+1},\cdots,E_N) \\
    &= \sum_{n=0}^k (-1)^n E_\alpha^n e_{k-n}(E_1, \cdots, E_N).
\end{aligned}
\end{equation}
It is defined that $e_0=1$ for either $N$ or $N-1$ variables. 
Taking the above equation into Eq. (\ref{Ka0}), we have
\begin{equation}
   \hat{K}_{\alpha} 
    =\sum_{n=0}^{N-1} \sum_{k=0}^{n} (-1)^{n+k} e_{n-k} E_{\alpha}^k H^{N-1-n},
    \label{Ka}
\end{equation}
where $e_k = e_k(E_1, \cdots, E_N)$ from now on. We then apply the Newton's identity
\begin{equation}
    e_k = \frac{1}{k} \sum_{n=1}^{k} (-1)^{n+1} p_n e_{k-n} , \label{ek}
\end{equation}
where 
\begin{equation}
    p_n= \trace (H^n)  =\sum_{\beta=1}^{N} (E_{\beta})^n
\end{equation}
is the $n$-th power sum. Via Eq. (\ref{ek}), we can, starting with $e_0=1$, iteratively change $e_k$ to a linear combination of the trace of a polynomial of the Hamiltonian. 

Here we provide another formula, which is more straightforward. By observation, 
\begin{equation}
    e_k' = \frac{1}{k} \sum_{n=1}^{k} (-1)^{n+1}  (p_n-E_\alpha^n) e_{k-n}', 
    \label{ekp}
\end{equation}
where $e_k'=e_k(E_1, \cdots, E_{\alpha-1},
E_{\alpha+1},\cdots,E_N)$. Similarly by iteration with $e_0'=1$, we can determine $e_k'$ for $k=0,\cdots,N-1$, and then obtain
\begin{equation}
   \hat{K}_{\alpha} =\sum_{k=0}^{N-1} (-1)^k e_k' H^{N-1-k}.
 \label{Ka2}
\end{equation}
Since $\trace \hat{P}_\alpha = 1,$ the denominator in Eq. (\ref{Pa}) is simply $\trace \hat{K}_\alpha$ and hence
\begin{equation}
    \hat{P}_\alpha = \frac{\hat{K}_\alpha}{\trace \hat{K}_\alpha}.
\end{equation}

Lastly, we show the formulae of first- and second-order derivatives of the eigenprojector. The process goes as follows:
\begin{equation}
    \partial_{\mu}\hat{P}_\alpha= \frac{1}{\trace \hat{K}_\alpha} \left[ \partial_{\mu}\hat{K}_\alpha - \trace(\partial_{\mu}\hat{K}_\alpha)\hat{P}_\alpha\right],
\end{equation}
and 
\begin{equation}
\begin{aligned}
    \partial_{\mu}\partial_{\nu}\hat{P}_\alpha =& \frac{1}{\trace \hat{K}_\alpha} \left\{ \partial_{\mu}\partial_{\nu}\hat{K}_\alpha 
    - \trace(\partial_{\mu}\partial_{\nu}\hat{K}_\alpha)\hat{P}_\alpha \right. \\
     &\left.-\trace(\partial_{\mu}\hat{K}_\alpha)\partial_{\nu}\hat{P}_\alpha 
    - \trace(\partial_{\nu}\hat{K}_\alpha)\partial_{\mu}\hat{P}_\alpha  \right\},
\end{aligned}
\end{equation}
where
\begin{equation}
    \partial_{\mu}\hat{K}_\alpha =\sum_{k=0}^{N-1} (-1)^k\left[ (\partial_{\mu}e_k') H^{N-1-k} + e_k' (\partial_{\mu}H^{N-1-k})\right],
\end{equation}
and
\begin{equation}
\begin{aligned}
    \partial_{\mu}\partial_{\nu}\hat{K}_\alpha =& \sum_{k=0}^{N-1} (-1)^k \left\{ (\partial_{\mu}\partial_{\nu}e_k' ) H^{N-1-k} 
    + e_k'(\partial_{\mu}\partial_{\nu} H^{N-1-k}) \right. \\
     &\left.+ (\partial_{\mu}e_k')\partial_{\nu}(H^{N-1-k})  
    + (\partial_{\nu}e_k')\partial_{\mu}(H^{N-1-k})  \right\}.
\end{aligned}
\end{equation}
The derivative of $e_k'$ is straightforward on Eq. (\ref{ekp}) with $\partial_\mu e_0'=0$, giving the iteration equation:
\begin{equation}
\begin{aligned}
    \partial_\mu e_k' = \frac{1}{k} \sum_{n=1}^{k} (-1)^{n+1} 
   & \left\{  ( \partial_\mu p_n- \partial_\mu E_\alpha^n) e_{k-n}' \right. \\
    & \quad \left. +(p_n-E_\alpha^n) \partial_\mu e_{k-n}'\right\} , 
\end{aligned}    
\end{equation}
and its second-order derivative is obtained consequently.
\begin{equation}
\begin{aligned}
    \partial_{\mu}\partial_{\nu}e_k' &= \frac{1}{k}\sum_{n=0}^{k} (-1)^{n+1} \left\{ 
    (\partial_\mu \partial_{\nu} p_n- \partial_\mu \partial_{\nu} E_\alpha^n) e_{k-n}'
    \right. \\
    & + (\partial_\mu  p_n- \partial_\mu E_\alpha^n) \partial_{\nu}e_{k-n}' 
    + ( \partial_{\nu} p_n-  \partial_{\nu} E_\alpha^n)  \partial_\mu e_{k-n}' \\
     &\left.+ (p_n-E_\alpha^n) \partial_\mu \partial_{\nu} e_{k-n}'  \right\}.
\end{aligned}
\end{equation}
The derivatives involve $E_{\alpha}^n$, which derivatives are now feasible through $E_{\alpha}^n = \trace \left(H^n \hat{P}_{\alpha} \right)$.
The participating derivatives of powers are 
\begin{equation}
    \partial_{\mu}  H^k = \sum_{n=1}^k H^{n-1}(\partial_\mu H)H^{k-n},
\end{equation}
\begin{equation}
\begin{aligned}
    \partial_{\mu} \partial_\nu H^k = \sum_{n=1}^k & \left\{ H^{n-1}(\partial_\mu \partial_\nu H)H^{k-n} 
     \right. \\   
    &+ (\partial_\mu H^{n-1})(\partial_\nu H)H^{k-n}    \\
    &\left. + H^{n-1}(\partial_\nu H) (\partial_\mu H^{k-n}) \right\},
\end{aligned}
\end{equation}
\begin{align}
    \partial_{\mu}  p_k = \trace(\partial_\mu H^k), \\
    \partial_{\mu} \partial_{\nu} p_k = \trace(\partial_\mu \partial_\nu H^k),
\end{align}
\begin{equation}
    \partial_{\mu}  E_\alpha^n = \trace \left[ (\partial_\mu H^n)\hat{P}_\alpha 
    + H^n(\partial_\mu \hat{P}_\alpha) \right],
\end{equation}
and
\begin{equation}
    \partial_{\mu} \partial_\nu  E_\alpha^n = \trace \left[ (\partial_\mu \partial_\mu H^n)\hat{P}_\alpha + (\partial_\mu H^n)(\partial_\nu \hat{P}_\alpha) \right].
\end{equation}

With $\partial_{\mu}\hat{P}_\alpha$ and $\partial_{\mu}\partial_{\nu}\hat{P}_\alpha$, the metric tensor $g_{\mu \nu}$ in Eq. (\ref{metric}), Christoffel symbols $\Gamma_{\alpha \beta \gamma}$ in Eq. (\ref{christoffel}), and hence the Riemann curvature $R_{\rho \lambda\mu\nu}$ in Eq. (\ref{riemann}) can be generated consequently.

\subsection{Multiple bands}
When multi-bands are considered, the multi-band projector and its derivatives are the sum of single-band ones, such as $\hat{P}=\sum_{\alpha\in \mathrm{val.}} \hat{P}_\alpha$. 
A special treatment occurs when energy-degeneracy happens at some $\vb{k}$, at which the formulas fail with the appearance of divergence. We need to reformulate for these special points. 

Let ${\{\alpha\}}$ denote the set of the indices of the degenerate bands with energy $E_{\alpha}$, and the number of degenerate bands be $d$. The corresponding eigenprojector would change to
\begin{equation}
    \hat{P}_{\{\alpha\}} = \prod_{\beta \neq \{\alpha\}} \left(\frac{H-E_\beta 1_{N}}{E_{\alpha} - E_{\beta}} \right). 
    \label{Pa2}
\end{equation}
The strategy is the same as before; we conquer the numerator, which is 
\begin{equation}
\begin{aligned}
   \hat{K}_{\{\alpha\}} &=\prod_{\beta \neq {\{\alpha\}}}  \left( H-E_\beta 1_{N} \right) \\
    &=\sum_{k=0}^{N-d} (-1)^k e_k''H^{N-d-k}.
\end{aligned}    \label{Kap}
\end{equation}
The main differences are: first, the maximal power of $H$ is $N-d$ since there are $d$ states exclusive from the product, and second, the elementary symmetric polynomials exclude $d$ variables in the set $\{E_{\alpha}\}$, which we define $e_k''=e_k(E_1, \cdots,\widecheck{\{E_{\alpha}\}},\cdots, E_{N})$ for brevity, where the entries below $~\check{}~$ have been omitted. The iteration equation for $e_k''$ is similar to Eq. (\ref{ekp}) with the aid of the Newton's identity, simply giving
\begin{equation}
    e_k'' = \frac{1}{k} \sum_{n=1}^{k} (-1)^{n+1}  (p_n-d E_\alpha^n) e_{k-n}'', 
    \label{ekpp}
\end{equation}
starting with $e_0''=1$.
After solving $e_k''$ and then $\hat{K}_{\{\alpha\}}$, the eigenprojector is
\begin{equation}
    \hat{P}_{\{\alpha\}} = \frac{d}{\trace \hat{K}_{\{\alpha\}}} \hat{K}_{\{\alpha\}},
\end{equation}
where the appearance of the factor $d$ is due to $\trace \hat{P}_{\{\alpha\}}  = d$. The following derivatives are closely the same as before, because $d E_{\alpha}^n$ in Eq. (\ref{ekpp}) will be taken as $d E_{\alpha}^n = \trace \left(H^n \hat{P}_{\alpha} \right)$. What would change from before are the factor $d$ in $\hat{P}_{\{\alpha\}}$ and their derivatives, and some upper limits of corresponding summations from $N-1$ to $N-d$.

\section{Construct normal basis matrices}
In this section, we outline a process of constructing the normal basis matrices, a set of basis matrices that are orthogonal to the tangents of the eigenprojector $\hat{P_\alpha}$. Given two tangents, $h_1 =\partial_{x}\hat{P_\alpha}$ and $h_2=\partial_{y}\hat{P_\alpha}$, the normal basis matrices are the complements of the tangents in the space. Since these two tangents are Hermitian as the eigenprojector, it is reasonable to prepare a complete set of Hermitian matrices. For the space of $N\times N$ Hermitian matrices, there are $N^2$ basis matrices: $H_1, \cdots, H_{N^2}$. We will make the basis matrices orthonormal: $\trace (H_{j} H_{k}) = \delta_{j k}$. To prepare a complete set of linearly independent matrices, a standard frame $\{M_j\}_{j=1}^{N^2}$ (unnormalized) is given as follows:
\begin{enumerate}
  \item The identity matrix,  $1_{N}$.
  \item $N(N-1)/2$ symmetric real matrices: matrix elements $M_{ab} = M_{ba}=1$ for $1\le a<b \le N$ and $0$ otherwise.
  \item $N(N-1)/2$ antisymmetric imaginary matrices: matrix elements $M_{ab} = M_{ba}=-i$ for $1\le a<b \le N$ and $0$ otherwise.
  \item $N-1$ diagonal traceless matrices: $M=\mathrm{diag}(1,\cdots,1,-j,0,\cdots,0)$, where the first $j$ elements are $1$ and the $(j+1)$-th is $-j$ for $j$ running from $1$ to $N-1$. 
\end{enumerate}
This recipe can produce the Pauli matrices for $N=2$ and the (scaled) Gell-mann matrices for $N=3$.

Then, we generate the basis matrices that construct the tangent plane. The Gram-Schmidt process is used to turn $h_1$ and $h_2$ into $H_1$ and $H_2$:
\begin{enumerate}
  \item Normalize $h_1~(\neq 0_N)$: $H_1 = h_1/\sqrt{\trace \left( h_1^2 \right)} $
  \item Orthogonalize $h_2$ against $H_1$: $H_2' = h_2 - \trace\left(H_1 h_2 \right) H_1$
  \item Normalize $H_2$: $H_2 = H_2'/\sqrt{\trace  (H_2')^2 } $
\end{enumerate}
The remaining basis matrices that are orthogonal to $H_1$ and $H_2$ will be constructed from the set of $M$'s through the Gram-Schmidt orthogonality process again: 
\begin{equation}
\begin{aligned}
    h_{j+2} &= M_j - \sum_{k=1}^{j-1} \trace \left( H_{k} M_j\right) H_k \quad \mathrm{and} \quad \\
    H_{j+2} &= h_{j+2}/\sqrt{\trace (h_{j+2})^2}
\end{aligned}
\end{equation}
for $j=1,\cdots, N^2-2$. (A check for whether $H_1$ and $H_2$ are linearly independent of $M_1$ and  $M_2$ is required. If dependence happens, more terms from $M$'s should be taken.) The normalized basis matrices, $H_3,\cdots, H_{N^2}$, will represent $\vec{n}^{\sigma'}$, which is identical to  $\vec{n}_{\sigma'}$, in Eq. (\ref{R1212}).

We remark that there are singular points at which the rank of the metric tensor is 1, at which points, $h_1$ and $h_2$ are linearly dependent. At those points, there will be $N^2-1$ normal basis matrices. 

\section{Completeness relation for basis matrices}
In this section, we show the proof of the completeness relation for basis matrices in Eq. (\ref{completeness}) and the formula of the Riemann curvature in Eq. (\ref{Riemann_curv}).

Suppose that $\{ H_j \}_{j=1}^{N^2}$ is a complete set of orthonormal Hermitian matrices of size $N\times N$: $\trace \left( H_j H_k\right) = \delta_{jk}$. For any arbitrary Hermitian matrix $M$, it can be uniquely expressed as $M = \sum_{j=1}^{N^2} M_j H_j$, where $M_j = \trace \left( H_j M\right)$. The $(a,b)$ entry of the matrix $M$ is 
\begin{equation}
\begin{aligned}
    M_{ab} =& \sum_{j=1}^{N^2} M_j (H_j)_{ab} \\
    = &  \sum_{j=1}^{N^2}\trace \left( H_j M\right) (H_j)_{ab} \\
    = &  \sum_{j=1}^{N^2} \sum_{c,d=1}^N M_{dc} \left( H_j \right)_{cd} (H_j)_{ab}. 
\end{aligned}
\end{equation}
This leads to the completeness relation (or the trace identity):
\begin{equation}
    \sum_{j=1}^{N^2} \left( H_j\right)_{ab} \left( H_j\right)_{cd} 
    = \delta_{ad} \delta_{bc}.
    \label{complete2}
\end{equation}
The completeness relation works for any conjugation: $H_j \mapsto H_j' = U H_j U^\dagger$, where $U\in \mathrm{U(N)}$, leading to
\begin{equation}
\begin{aligned}
     \sum_{j} & \left( H_j'\right)_{ab} \left( H_j'\right)_{cd} = \sum_{j} \left( U H_j U^\dagger \right)_{ab} \left( U H_j U^\dagger \right)_{cd} \\
     & = \sum_j \sum_{m,n,s,t} U_{am} (H_j)_{mn} (U^\dagger)_{nb} 
     U_{cs} (H_j)_{st} (U^\dagger)_{td} \\
     & = \sum_{m,n,s,t} \delta_{mt} \delta_{ns} U_{am} (U^\dagger)_{nb} U_{cs} (U^\dagger)_{td} \\
     & = \sum_{m,n} U_{am}(U^\dagger)_{md} U_{cn} (U^\dagger)_{nb} \\
     & = \delta_{ad} \delta_{bc}.
\end{aligned}
\end{equation}

Therefore, the term in Eq. (\ref{Riemann_1}) becomes

\begin{equation}
\begin{aligned}
    \partial_{\mu} \vec{e}_\rho & \cdot \partial_{\nu} \vec{e}_\lambda
    = \trace\left[ \left(  \partial_{\mu}  \partial_{\rho} \hat{P}_{\alpha}\right) 
    \left( \partial_{\lambda}  \partial_{\nu} \hat{P}_{\alpha}\right) \right] \\
    & = \sum_{a,b,c,d} \delta_{ad} \delta_{bc}  \left(  \partial_{\mu}  \partial_{\rho} \hat{P}_{\alpha}\right)_{ab}   
    \left( \partial_{\lambda}  \partial_{\nu} \hat{P}_{\alpha}\right)_{cd}  \\
    & = \sum_j \sum_{a,b,c,d}  \left(  \partial_{\mu}  \partial_{\rho} \hat{P}_{\alpha}\right)_{ab}  (H_j)_{ba} (H_j)_{dc}  \left( \partial_{\lambda}  \partial_{\nu} \hat{P}_{\alpha}\right)_{cd} \\
    & = \sum_j \trace\left[ \left(  \partial_{\mu}  \partial_{\rho} \hat{P}_{\alpha}\right) 
    H_j \right]
    \trace\left[H_j
    \left( \partial_{\lambda}  \partial_{\nu} \hat{P}_{\alpha}\right) \right] \\
    &= \left(   \partial_\mu \vec{e}_{\lambda} \cdot \vec{e}^\sigma \right)
    \left( \vec{e}_\sigma \cdot \partial_\nu \vec{e}_{\rho} \right) 
    +\left( \partial_\mu \vec{e}_{\lambda} \cdot \vec{n}^{\sigma'} \right)
    \left( \vec{n}_{\sigma' }\cdot \partial_\nu \vec{e}_{\rho} \right).
\end{aligned}
\end{equation}
In the above third equality, we apply Eq. (\ref{complete2}) of exchanging $a$ with $b$ and $c$ with $d$. 
For a normal basis matrix $H_j$ in the orthonormal set, it is identical to $\vec{n}_j$ as well as $\vec{n}^j$, so $\sum_j H_j \otimes H_j = \sum_j \vec{n}^j \otimes \vec{n}_j$. 
For the tangents $h_\mu :=\partial_{\mu}\hat{P_\alpha}~ (\mu=1,2)$, their dual matrices are $h^{\mu} = g^{\mu \nu} h_{\nu}$. We assume a linear transformation between $h_{1,2}$ and $H_{1,2}$ to be $H_\mu = \sum_{\nu=1,2} f_{\mu \nu}h_\nu$ and its inverse, which exists when $g$ is invertible, $h_\mu = \sum_\nu (f^{-1})_{\mu \nu} H_\nu$. The coefficients $f$'s and the metric tensor are related since 
\begin{equation}
\begin{aligned}
    g_{\mu \nu} &= \trace (h_\mu h_\nu) \\
    &= \sum_{\rho \sigma} (f^{-1})_{\mu \rho} (f^{-1})_{\nu \sigma} \trace(H_\rho H_\sigma) \\
    &=\sum_\rho (f^{-1})_{\mu \rho} (f^{-1})_{\nu \rho}
\end{aligned}
\end{equation}
Its inverse gives $g^{\mu \nu} = \sum_\rho f_{\rho \mu} f_{\rho \nu}$. Finally, it is evident that
\begin{equation}
\begin{aligned}
    \vec{e}^{\mu} \otimes \vec{e}_{\mu} & \dot = \sum_\mu h^\mu \otimes h_\mu \\
    &= \sum_{\mu \rho} g^{\mu \rho} h_\rho \otimes h_\mu \\
    &= \sum_{\mu \sigma} f_{\sigma \mu} f_{\sigma \rho} h_\rho \otimes h_\mu \\
    &= \sum_\sigma H_\sigma \otimes H_\sigma .
\end{aligned}
\end{equation}

\section{singular curve and adapted coordinates} \label{appendix:apated_coor}
The kernel vector $(\eta^1,\eta^2)^\mathsf{T}$  is determined by 
$\partial_{\eta}\hat{P}_\alpha=\eta^{\mu}\partial_{\mu}\hat{P}_\alpha=0$. In the two-band model, the equation becomes to the vector equation:
\begin{equation}
    \partial_{\eta} \vec{d} = \frac{1}{2d^2} \vec{d}\partial_{\eta}d^2.
\end{equation}
Using $\vec{d}$ in Eq. (\ref{eq:d_vec}), we have
\begin{equation}
    \frac{\eta^1 \cos k_x}{\sin k_x} = \frac{\eta^2 \cos k_y}{\sin k_y} = \frac{\eta^1 \sin k_x+\eta^2 \sin k_y}{m_0 - \cos k_x - \cos k_y}.
\end{equation}
The equation for the kernel vector is
\begin{equation}
    \frac{\eta^2}{\eta^1} = \frac{\tan k_y}{\tan k_x}, \label{v0}
\end{equation}
and then the equation for the singular points is
\begin{equation}
    \begin{aligned}
    F(k_x,k_y) =& \tan k_x \sin k_x + \tan k_y \sin k_y + \cos k_x +\cos k_y \\
    &- m_0=0.
    \end{aligned}
\end{equation}
The set of the singular points is be a continuous closed curve surrounding $(\pi,\pi)$, and along the curve the kernel field rotates one round as shown in Fig.~\ref{fig:sing_pts}(a). The singular curve passes through the peak points at $(k_x,k_y) = \pm (\pi/2,\pi/2)$ and $\pm (\pi/2,-\pi/2)$. At these special points, the kernel field is tangent to the curve, giving $\eta_1=-\eta_2$ and $\eta_1=\eta_2$, respectively. The images at these points are cusp points [see Fig.~\ref{fig:sing_pts}(b)]; at a cusp the change of the orientation of the tangent changes by an external angle, $\abs{\theta} = \pi$.

\begin{figure}[tb]
\centering
\includegraphics[width=0.45\textwidth]{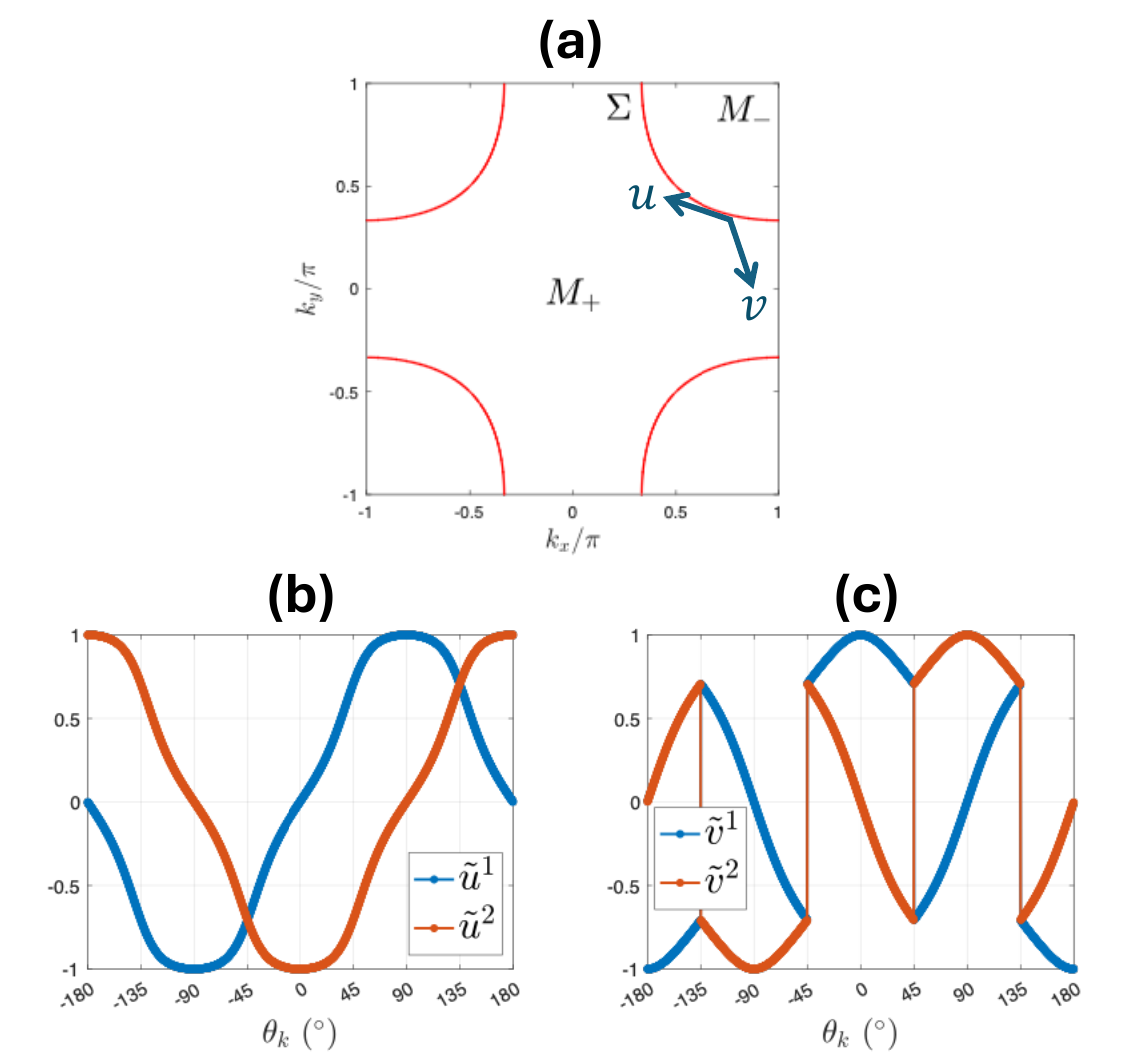}
\caption{(a) The adapted coordinates $(u,v)$ along the singular curve. (b) The normalized tangent vectors $(\tilde{u}^1,\tilde{u}^2)$ from Eq. (\ref{u_cho}), and (c) the normalized kernel vector $(\tilde{v}^1,\tilde{v}^2)$ from Eq. (\ref{v_cho}) in the singular curve. The angle $\theta_k$ is relative to the $k_x$-axis around $(\pi,\pi)$. Our choice has continuous $u$'s but has $u$'s discontinuous at the preimages of the cusps. This choice can produce a consistent positive frame: $\mathrm{sgn}(u\wedge v)>0$.}
\label{fig:uv_norm}
\end{figure}

The tangent field to the singular curve is obtained through $X_u F=u^\mu\partial_\mu F=0$, giving
\begin{equation}
    \frac{u^2}{u^1} = -\frac{\cos^2 k_y \sin k_x}{\cos^2 k_x \sin k_y}. 
    \label{u0}
\end{equation}

In the singular curve, an adapted coordinate system $(u,v)$ is chosen to simplify our problem. The $u$-direction is along the singular curve and the $v$-direction is the direction of the kernel field $\eta$. In the calculating Christoffel symbols,
the directional derivatives $\partial_u = \frac{\partial k^\mu}{\partial u} \partial_{\mu}:=u^\mu \partial_{\mu}$ and $\partial_v = \frac{\partial k^\mu}{\partial v} \partial_{\mu}:=v^\mu \partial_{\mu}$ are required. Perceptibly, it is more convenient to regard $u$ and $v$ as independent variables, such that the derivatives are commutable: $\partial_{u} \partial_{v} = \partial_{v} \partial_{u}$. This is the Clairaut-Schwarz theorem. However, it is found that an arbitrary choice of $u^\mu$ and $v^\mu$ that satisfy the ratios in Eqs. (\ref{u0}) and (\ref{v0}) cannot make them independent. 

To resolve this problem, we look at the inverse of the Jacobian $\left( \frac{\partial k^{\mu}}{\partial z^a}\right)^{-1}=\left( \frac{\partial z^a}{\partial k^{\mu}}\right) $, where $z^{1,2}= u,v$, giving
\begin{equation}
\begin{aligned}
    \mqty(\frac{\partial u}{\partial k_x} & \frac{\partial u}{\partial k_y} \\ \frac{\partial v}{\partial k_x} & \frac{\partial v}{\partial k_y}) 
    &=\mqty(u^1 & v^1 \\ u^2 & v^2)^{-1} 
      = \frac{1}{u\wedge v} \mqty(v^2 & -v^1 \\ -u^2 & u^1) \\ &=:\mqty(V^2 & -V^1 \\ -U^2 & U^1) .
\end{aligned}
\end{equation}
Assume that $u\wedge v=u^1v^2 - u^2 v^1 \neq 0$ for linearly independent vectors. We look for $\vec{u},\vec{v}$ that can satisfy the commutable equations $\frac{\partial}{\partial k_x} \frac{\partial z^a}{\partial k_y} = \frac{\partial}{\partial k_y} \frac{\partial z^a}{\partial k_x}$, which equivalently are
\begin{equation}
    \frac{\partial}{\partial k_x} W^1 = - \frac{\partial}{\partial k_y} W^2,
\end{equation}
where $W=U$ or $V$. A simple choice is to make the equation be zero. We choose 
\begin{equation}
    U^1 = \frac{\sin k_y}{\cos^2 k_y} \quad \mathrm{ and } \quad
    U^2 = -\frac{\sin k_x}{\cos^2 k_x} ,
\end{equation}
and
\begin{equation}
    V^1 = \frac{1}{\tan k_y}  \quad \mathrm{ and } \quad
    V^2 = \frac{1}{\tan k_x}.
\end{equation}
Inverse of the Jacabian suggests us the choice of components to be
\begin{equation}
    \begin{aligned}
    \mqty(u^1 \\ u^2) = \frac{\abs{\sin k_x \sin k_y}}{D}
    \mqty(\cos^2 k_x \sin k_y \\ -\cos^2 k_y \sin k_x ) , 
    \end{aligned}
    \label{u_cho}
\end{equation}
and
\begin{equation}
    \begin{aligned}
    \mqty(v^1 \\ v^2) = -\frac{\cos^3 k_x \cos^3 k_y}{D}\mqty(\tan k_x \\ \tan k_y) , 
    \end{aligned}
    \label{v_cho}
\end{equation}
where $D=\cos^3 k_x \sin^2 k_y + \cos^3 k_y \sin^2 k_x$. The sign factor in $u^{1,2}$ is intentionally introduced to make a positive frame and continuous $u$'s. The normalized adapted $u^\mu$ and $v^\mu$ around the singular curve are plotted in Fig.~\ref{fig:uv_norm}. However, our solution produces divergence and discontinuity in $\vec{v}$ at $\pm (\pi/2.\pm \pi/2)$, where the singular curvature is also divergent.



\bibliographystyle{apsrev4-1}
\bibliography{GB}

\end{document}